\documentclass[prb,aps,twocolumn,groupedaddress,showpacs,floatfix]{revtex4}

\usepackage{graphicx}
\usepackage{amsmath}

\begin{document}
\bibliographystyle{apsrev}

\title{Transport Properties of a spinon Fermi surface coupled to a U(1) gauge field}

\author{Cody P. Nave}
\author{Patrick A. Lee}
\affiliation{Department of Physics, Massachusetts Institute of Technology,
Cambridge, Massachusetts 02139}
\date{\today}

\begin{abstract}
With the organic compound $\kappa$-(BEDT-TTF)$_2$-Cu$_2$(CN)$_3$ in mind,
we consider a spin liquid system where a spinon Fermi surface is coupled 
to a U(1) gauge field.  Using the non-equilibrium Green's function formalism,
we derive the Quantum Boltzmann Equation (QBE) for this system.  In this system, 
however, one cannot \textit{a priori} assume the existence of Landau
quasiparticles.  We show that even without this assumption one can
still derive a linearized equation for a generalized distribution function.  
We show that the divergence of the effective mass and of the finite 
temperature self-energy do not enter these transport coefficients and thus
they are well-defined. Moreover, using a variational method, we calculate the temperature
dependence of the spin resistivity and thermal conductivity of this system.

\end{abstract}


\maketitle

\section{Introduction}

Recent experiments have shown evidence that the organic compound 
$\kappa$-(BEDT-TTF)$_2$-Cu$_2$(CN)$_3$ maybe the first experimentally realized
spin liquid in dimension greater than one.\cite{Shimizu:03} This quasi two-dimensional
material can be described as an effectively isotropic
spin $1/2$ system on a triangular lattice.  It is found experimentally
to be insulating and shows no evidence of long range magnetic order down 
to mK temperatures.  Fitting the susceptibility using the
high temperature series expansion of the spin $1/2$ Heisenberg model on a 
triangular lattice, the exchange coupling $J$ is roughly $250K$.
The static spin susceptibility also remains finite down to the lowest temperatures 
measured.\cite{Shimizu:03}
Because of the lack of magnetic order even at temperatures many orders of magnitude 
lower than the exchange coupling $J$ and the experimental evidence for abundant low
energy spin excitations,
there has been a proposal that this system may be well described by a spin-liquid where a
spinon Fermi surface is coupled to a $U(1)$ gauge field.\cite{SSLee:05,Motrunich:05}
Recent work with this model has lead to the suggestion of a possible spinon pairing state
that may explain the observed features in the experimental measurements of the 
specific heat and magnetic
susceptibility.\cite{SSLee:07, Nave:07}   In this paper, we focus on the original
model, \textit{i.e.} we consider temperatures above the pairing transition temperature but
small compared to the exchange temperature.  In this regime, we start with our model
Lagrangian and proceed to develop a version of the quantum Boltzmann equation (QBE).
Despite potential pitfalls that we discuss below, we show that the QBE is well-defined
and apply it to systems with steady-state thermal and spin currents.  We show that
the transport coefficients are finite and in particular calculate the 
temperature dependence of the spin resistivity and the more experimentally accessible thermal
conductivity.

We begin by considering the $t$--$J$ model on the triangular lattice.  We construct a mean-field
state by applying the slave-boson
formalism and enforcing the local constraint of no double occupancy exactly.  It is known that
considering fluctuations around this mean-field state leads to a $U(1)$ gauge theory.\cite{Lee:92}
More recently, the slave-rotor representation has been applied to the Hubbard model on a triangular
lattice and it was shown that again one can arrive at a $U(1)$ gauge theory.\cite{Florens:04,SSLee:05}
Because of the large number of low energy excitations due to presence of the spinon Fermi surface, 
we assume that a deconfined state occurs and thus we consider a non-compact $U(1)$ gauge
theory.  In other words, we assume that instanton effects are negligible. Thus our starting point
is the Lagrangian for a 2-D spinon Fermi surface system coupled to a non-compact $U(1)$ 
gauge field given by 
\begin{equation}
\mathcal{L} = \psi^\ast_\sigma \left( \partial_0 - i a_0 - \mu \right) \psi_\sigma +
\frac{1}{2 m}\psi^\ast_\sigma \left(-i \nabla - \mathbf{a}\right)^2 \psi_\sigma,
\label{lagrangian}
\end{equation}
where the gauge field kinetic energy term has been dropped because its strength is
inversely proportional to the charge gap which is large since we are assuming we are in
the insulating phase. $\psi_\sigma$ is the spinon field and the
gauge field is $a = (a_0, \mathbf{a})$.  $\mu$ is the chemical potential.  We work
in Coulomb gauge $\nabla \cdot \mathbf{a} = 0$.  

We can then proceed to integrate out the spinons in order to generate dynamics
for the gauge field.  This cannot be done exactly; however, we can work in the 
Gaussian approximation also known as the random-phase approximation (RPA).  We
then consider spinons coupled to the effective action for the gauge field coming
from the spinon bubbles. The use of the RPA can be formally arranged in 
the standard way through the $1/N$ expansion by introducing $N$ species of fermions.\cite{Polchinski:94}

The longitudinal part of the gauge propagator is related to the
density-density response and thus does not show any singular behavior 
for low energies and momentum.  The transverse part however does give
rise to long range interactions. After integration, we find that the 
effective action for the gauge field is $S(a) = \sum_q \Pi(q) a^\dagger_q a_q$, where 
\begin{equation}
\Pi(q) = \frac{\gamma v_\mathrm{F} \left|q_0\right|}{\sqrt{v^2_\mathrm{F} \mathbf{q}^2 +q^2_0}
+ q_0 } + \chi_\mathrm{D} \mathbf{q}^2,
\end{equation}
where $\chi_\mathrm{D} = \frac{1}{12 \pi m}$ and $\gamma = \frac{k_\mathrm{F}}{\pi}$.
Thus the effective gauge field propagator is given by $D(q) = \Pi(q)^{-1}$.  
Rotating back to real time, $q_0 = i \nu$, and working in the limit that $\nu \ll v_\mathrm{F} q$,
the gauge propagator becomes
\begin{equation}
D(\mathbf{q}, \nu) = \frac{1}{-i \gamma \frac{\nu}{q} + \chi_\mathrm{D} q^2},
\label{eq:prop}
\end{equation}
where q is now the magnitude of $\mathbf{q}$.\cite{Ioffe:89,Lee:92}

For the remainder of the paper we consider the effective theory given by taking the Lagrangian of 
Eq. \ref{lagrangian} and adding gauge field dynamics through the RPA propagator of Eq. \ref{eq:prop}.
This particular gauge theory has been studied previously in the context of the half-filled
Landau level (Ref. \onlinecite{Halperin:93}) and high-temperature superconductors 
(Refs. \onlinecite{Lee:92} and \onlinecite{Nagaosa:90}.)  In particular, the 
spinon self-energy correction due to the RPA gauge propagator has been examined.
Again one finds that the most singular correction comes from considering the transverse
gauge field fluctuations.  To one-loop order, the self-energy $\Sigma(\mathbf{k},\omega)$ 
is found to be $\mathrm{Re} \Sigma \sim \mathrm{Im} \Sigma \sim \omega^{2/3}$.   
We note that this implies a vanishing quasiparticle spectral weight, \textit{i.e.}
the Landau criterion for quasiparticles is violated and thus the Fermi liquid picture
is invalid for this system.  Moreover, the effective mass is found to diverge at the
Fermi surface as $\xi_\mathbf{k}^{-1/3}$ where $\xi_\mathbf{k} = \epsilon_\mathbf{k} -\mu
= k^2/2m-\mu$.\cite{Halperin:93}

Despite the fact that quasiparticles are ill-defined in this system, we examine
the standard expressions for the spin resistivity and thermal conductivity.
The spin resistivity is given by $\rho_S = \frac{m}{n \tau}$
where $1/\tau$ is the momentum relaxation rate.  From the self-energy correction to the
fermion propagator, we calculate the momentum relaxation rate 
$1/\tau \sim T^{-4/3}$.\cite{Lee:92}  Beyond the assumption of the validity of the
quasiparticle picture, in order to arrive at $\rho_S$, we also need to consider the
effective rather than bare mass for the spinons; however, as mentioned above the
effective mass is divergent. In section IV, we find that $\rho_S \sim T^{4/3}$, a
result which is consistent with the calculation when the quasiparticles are assumed 
to be well-defined and the renormalization of the mass is ignored.

The standard simple result for the thermal conductivity gives $\kappa \sim C v^2 \tau_E$ 
where $C$ is the specific
heat, $v$ is the particle velocity and $1/\tau_E$ is the energy relaxation rate.
Again from the self-energy correction to the spinon propagator, we calculate the
energy relaxation rate $1/\tau_E \sim T^{2/3}$.\cite{Lee:92}  For a system of 
fermions $C = \gamma T$ and the velocity would be temperature independent.  Thus
$\kappa/T \sim T^{-2/3}$.  However these assumptions are again not justified in view
of the divergent effective mass and lack of well-defined quasiparticles.  
Considering the mass renormalization leads to
a specific heat $C \sim T^{2/3}$ and thus $\kappa/T \sim T^{-1}$; moreover, 
the velocity goes to zero as the effective mass diverges.  Thus 
it is unclear how to proceed. In section V, we see that 
the power law $T$ dependence of the thermal conductivity given by the naive arguments ignoring the effects
of the effective mass turns out to be correct.

Because of these issues, we are forced to consider the interactions between the spinons and
gauge bosons more carefully and thus turn to a quantum Boltzmann description of the system.
As mentioned above, we cannot derive a QBE using the Landau quasiparticle picture because Fermi-liquid theory
is invalid for this system.  We find however that we can proceed by following the work of Prange and Kadanoff 
who studied the electron-phonon system at temperatures high compared to the Debye temperature. 
\cite{Prange:64}  At high temperatures, the electrons rapidly emit phonons so that their
precise energy is not well defined.  Thus they deal with an analogous situation where the
quasiparticle picture breaks down.  Closely following their work, we find that if 
the self-energy at small frequencies is independent of $\xi_\mathbf{k}$, we can define
a generalized distribution function and derive a closed equation describing the dynamics of this 
generalized distribution function.  This equation is analogous to the standard Fermi-liquid QBE. 
We note that the derivation of the QBE for this system is very closely related to the work done by
Kim \textit{et al.} in Ref. \onlinecite{Kim:95} studying the $\nu = 1/2$
fraction quantum Hall state, except that we derive the QBE in a different coordinate
system and linearize in a different way, which allows us to use variational methods to
calculate the transport properties.  Note that a similar derivation of the QBE for
generalized distribution functions is also considered in the work by Mahan.\cite{Mahan:83}

\section{Deriving the QBE}
To derive the QBE for this system, we work in the standard non-equilibrium
Green's function formulation. We begin with two matrices of Green's functions
$\tilde{G}$ and $\tilde{\Sigma}$ that satisfy Dyson's equation
\begin{equation}
\tilde{G} = \tilde{G_0} \tilde{\Sigma} \tilde{G},
\label{eq:Dyson}
\end{equation}
where
\begin{equation}
\tilde{G} = \left[ \begin{array}{ccc} G_t&-G^< \\ 
G^> & -G_{\bar{t}} \end{array} \right]
\end{equation}
with $\tilde{\Sigma}$ defined similarly.  Note that following Ref. \onlinecite{Mahan},
in Eq. \ref{eq:Dyson} the product of two functions actually implies an integration over
a shared space-time variable.  Also here we use the
standard definitions, following Refs. \onlinecite{Kadanoff} and \onlinecite{Mahan},
\begin{eqnarray}
\label{eq:Gdef}
G^>(x_1,x_2) &=& -i \left< \psi(x_1) \psi^\dagger(x_2) \right> \\
G^<(x_1,x_2) &=&  i \left< \psi^\dagger(x_2) \psi(x_1) \right> \\
G_t(x_1,x_2) &=& \Theta(t_1-t_2) G^>(x_1,x_2) + \nonumber \\
&&\Theta(t_2-t_1) G^<(x_1,x_2) \\
G_{\bar{t}}(x_1,x_2) &=& \Theta(t_2-t_1) G^>(x_1,x_2) + \nonumber \\
&&\Theta(t_1-t_2) G^<(x_1,x_2),
\end{eqnarray}
with associated self energies $\Sigma^>$, $\Sigma^<$, $\Sigma_t$ and
$\Sigma_{\bar{t}}$.  Here $x = (\mathbf{r},t)$. $G_0$ denotes the 
non-interacting Green's functions.
These Green's functions are related to the standard retarded ($G^\mathrm{R}$)
and advanced ($G^\mathrm{A}$) Green's functions through
\begin{eqnarray}
G^R &=& G_t - G^< = G^>-G_{\bar{t}} \\
G^A &=& G_t - G^> = G^<-G_{\bar{t}}.
\label{eq:GRGA}
\end{eqnarray}
We perform a change of variables so that all the Green's functions are
expressed in terms of relative and center of mass like coordinates.
Throughout this paper, we work with the Fourier transform of the
relative coordinates so that we can write $G^<(\mathbf{k},\omega,\mathbf{r},t)$.

For a general system of fermions in thermal equilibrium, we can write that
\begin{eqnarray}
G^<(\mathbf{k},\omega) &=& i f_0(\omega) A(\mathbf{k},\omega) \\
G^>(\mathbf{k},\omega) &=& - i (1 - f_0(\omega)) A(\mathbf{k},\omega)
\end{eqnarray}
where $f_0(\omega)$ is the Fermi distribution function at some temperature $T$.
$A(\mathbf{k},\omega) = - i (G^R(\mathbf{k},\omega)-G^A(\mathbf{k},\omega))$ is 
the spectral function given by
\begin{equation}
A(\mathbf{k},\omega) = \frac{-2 \mathrm{Im} \Sigma^R(\mathbf{k},\omega)}
{ \left[\omega - \xi_k - \mathrm{Re} \Sigma^R(\mathbf{k},\omega) \right]^2 +
\left( \mathrm{Im} \Sigma^R (\mathbf{k},\omega) \right)^2}.
\label{eq:fullA}
\end{equation}
\vspace{1mm}

In Fermi liquid theory, the quasiparticles are well defined because
Im $\Sigma^R \sim \omega^2 \ll \omega$ for small $\omega$.  This means that the equilibrium spectral function 
is sharply peaked as a function of $\omega$, so that ignoring the incoherent background, it can be written as
\begin{equation}
A(\mathbf{k},\omega) = 2 \pi \delta \left( \omega - \xi_\mathbf{k} - \mathrm{Re} \Sigma^R
(\mathbf{k},\omega) \right).
\label{eq:normalA}
\end{equation}
In this paper we are considering linear response, so we assume that the system
is slowly varying in space and time and that therefore there is a
notion of a local equilibrium temperature $T$ for every $(\mathbf{r},t)$.
Assuming that the system remains close enough to equilibrium that the sharp $\omega$ 
peaking of the spectral weight remains valid, the standard Landau quasiparticle 
QBE for the fermion distribution function $f(\mathbf{k},\mathbf{r},t)$ then follows. 

As mentioned in the Introduction, in this model which is described by Eqs. \ref{lagrangian} and 
\ref{eq:prop}, both the real and 
the imaginary parts of the self-energy  of the spinons scale as $\omega^{2/3}$ for 
small $\omega$. This violation of the Landau criterion for the existence of well-defined
quasiparticles invalidates the normal derivation of the QBE since
the spectral weight is no longer sharply peaked in $\omega$ and thus cannot be written 
in the form of Eq. \ref{eq:normalA}.

We find that we can still proceed to derive a QBE for this system due to the
form of the self-energy. First we change variables from $\mathbf{k}$ to 
$\xi \equiv \xi_\mathbf{k}$ and $\mathbf{\hat{k}}$.  Then we
note that since the self-energy is independent of the magnitude of $\mathbf{k}$, 
we can write $\Sigma^R(\mathbf{k},\omega) = \Sigma^R(\mathbf{\hat{k}},\omega)$, and that 
then for small enough $\omega$, $A(\mathbf{k},\omega)$ is a peaked function
of $\xi$ around $\xi=0$.  This property of the invariance of the self-energy with the magnitude of
$\mathbf{k}$, \textit{i.e.} that it is only a function of $\omega$ and $\mathbf{\hat{k}}$,
is exactly the same property that Prange and Kadanoff used in Ref. \onlinecite{Prange:64} 
to derive a generalized QBE for the electron-phonon system that is valid at temperatures
high relative to the Debye temperature.

Following the work of Ref. \onlinecite{Prange:64}, we assume that the system remains close enough 
to local equilibrium that the self-energy is independent of $\xi$ at all times.
Combining this assumption with the fact that the $\int \frac{d\xi}{2\pi} A = 1$, 
it follows that $G^<$ and $G^>$ are sharply peaked functions of $\xi$. 
Integrating over the region of peaking in $\xi$, 
we can then define the generalized distribution function 
$f(\mathbf{\hat{k}},\omega,\mathbf{r},t)$ as
\begin{equation}
\label{eq:gdf}
\int \frac{d\xi}{2\pi} \left[ -i\, G^<(\xi,\mathbf{\hat{k}},\omega,\mathbf{r},t)\right] = 
f(\mathbf{\hat{k}},\omega,\mathbf{r},t),
\end{equation}
\vspace{10mm}where $f(\mathbf{\hat{k}},\omega,\mathbf{r},t)$ is the density of spinons with momentum in the
$\mathbf{\hat{k}}$ direction, energy $\omega$ at a given position $\mathbf{r}$ and time $t$.
Similarly we have
\begin{equation}
\int \frac{d\xi}{2\pi} \left[ i\, G^>(\xi,\mathbf{\hat{k}},\omega,\mathbf{r},t)\right] = 
1-f(\mathbf{\hat{k}},\omega,\mathbf{r},t).
\end{equation}

Having established the definition of the generalized distribution function in a system
without well-defined Landau quasiparticles, we now proceed to derive the QBE that
governs this distribution. 
We begin with the matrix form of Dyson's equation (Eq. \ref{eq:Dyson}.)  In particular, we
need to derive the equation of motion for $G^<$.  After a standard short derivation, 
see for instance Refs. \onlinecite{Mahan} and \onlinecite{Kim:95}, and working 
in the gradient expansion, we arrive at the expression, 
\begin{align}
\nonumber
\left[\omega-\frac{k^2}{2m}-\mathrm{Re}\,\Sigma^\mathrm{R},G^<\right]-\left[\Sigma^<,
\mathrm{Re}\, G^\mathrm{R}\right] \\ 
=\Sigma^>G^<-G^>\Sigma^<,
\label{eq:QBE1}
\end{align}
which describes the evolution of $G^<$. Here $[A,B]$ is a generalized Poisson bracket defined as
\begin{align}
[A,B]=\frac{\partial A}{\partial \omega}\frac{\partial B}{\partial t} -
\frac{\partial A}{\partial t}\frac{\partial B}{\partial \omega} + 
\nabla_{r}A\cdot\nabla_{k}B-\nabla_{k}A\cdot\nabla_{r}B.
\end{align}
Note that in Eq. \ref{eq:QBE1}, we have suppressed the variables 
$(\mathbf{k},\omega,\mathbf{r},t)$ for the self-energies and Green's functions.

From standard perturbation theory working to one-loop order, the self-energies 
$\Sigma^<$ and $\Sigma^>$ are given by
\begin{widetext}
\begin{eqnarray}
\Sigma^<&=&\sum_\mathbf{q} \int_0^\infty \frac{d\nu}{\pi} \left|\frac{\mathbf{k}\times
\mathbf{\hat{q}}}{m}\right| ^2 \mathrm{Im} D(\mathbf{q},\nu) \left[(n_0(\nu)+1)
G^<(\mathbf{k}+\mathbf{q},\omega+\nu) + n_0(\nu)G^<(\mathbf{k}+\mathbf{q},\omega-\nu) \right]\\
\Sigma^>&=&\sum_\mathbf{q} \int_0^\infty \frac{d\nu}{\pi} \left|\frac{\mathbf{k}\times
\mathbf{\hat{q}}}{m}\right| ^2 \mathrm{Im} D(\mathbf{q},\nu) \left[n_0(\nu)
G^>(\mathbf{k}+\mathbf{q},\omega+\nu) + (n_0(\nu)+1)G^>(\mathbf{k}+\mathbf{q},\omega-\nu) \right],
\end{eqnarray}
\end{widetext}
where for notation convenience, we have dropped the variables $(\mathbf{r},t)$.
Note that we have assumed that the gauge bosons are always in local thermal equilibrium and that 
$n_0(\nu) = 1/(e^{\beta \nu}-1)$, the standard boson equilibrium distribution function at 
temperature $T$.  We therefore are studying the contributions to the transport coefficients
arising from spinons. In Section VI we study the validity of this assumption for the particular
case of the thermal conductivity.

To derive the QBE for the generalized distribution function defined in Eq. \ref{eq:gdf},
we need to write Eq. \ref{eq:QBE1} in terms of $f(\mathbf{\hat{k}},\omega,\mathbf{r},t)$.
We integrate both sides of Eq. \ref{eq:QBE1} over the magnitude $\xi$ and rely on the 
assumption of the peaking as a function of $\xi$.  
From the Kramer's Kroenig relation,
\begin{equation}
\mathrm{Re}\,G^R(\xi,\mathbf{\hat{k}},\omega) = 
- \mathcal{P} \int \frac{d \omega'}{\pi} 
\frac{\mathrm{Im}\,G^R(\xi,\mathbf{\hat{k}},\omega')}{\omega - \omega'}.
\end{equation}
Since $-2 \, \mathrm{Im}\, G^R = A$, the condition that $\int \frac{d\xi}{2\pi} A = 1$
implies that
\begin{equation}
\int \frac{d\xi}{2\pi} \mathrm{Re}\,G^\mathrm{R}
= \mathcal{P} \int \frac{d\omega'}{2\pi} \frac{1}{\omega - \omega'} = 0.
\end{equation}
Therefore with the assumption that $\Sigma^<$ is independent of $\xi$,
we can drop the second term on the LHS in Eq. \ref{eq:QBE1} and 
we are left with
\begin{align}
\nonumber
\int d\xi &\left[\omega-\frac{k^2}{2m}-\mathrm{Re}\,\Sigma^\mathrm{R},G^<\right] \\
&=\int d\xi \left( \Sigma^>G^<-G^>\Sigma^< \right).
\label{eq:QBE2}
\end{align}

We expand the remaining generalized Poisson bracket on the LHS and the
QBE becomes, 
\begin{equation}
\begin{split}
\int & d\xi  \biggl[ \left( 1 - \frac{\partial \,\mathrm{Re}\Sigma^R}{\partial \omega} \right) 
\frac{\partial G^<}{\partial t}  
+ \frac{\partial \,\mathrm{Re} \Sigma^R}{\partial t} \frac{\partial G^<}{\partial \omega}\,  - \\
&\nabla_\mathbf{r} \mathrm{Re} \Sigma^R \cdot \nabla_\mathbf{k} G^< +
\nabla_\mathbf{k} \left( \epsilon_k + \mathrm{Re} \Sigma^R \right) \cdot \nabla_\mathbf{r} G^<
\biggr] = I_\mathrm{coll},
\label{eq:fullQBEunint}
\end{split}
\end{equation}
where the collision integral, $I_\mathrm{coll}$, is defined below.  
Using our assumptions that the self-energies depend only on $\omega$ even when
the system is not in equilibrium and that $G^<$ remains a well-peaked function of
$\xi$, we can perform the integration over
$\xi$ and find that
\begin{equation}
\label{eq:fullQBE}
\begin{split}
&\left( 1 -  \frac{\partial \mathrm{Re} \Sigma^R}{\partial \omega} \right) 
\frac{\partial f}{\partial t}  
+ \frac{\partial \mathrm{Re} \Sigma^R}{\partial t} \frac{\partial f}{\partial \omega}\, -
\nabla_\mathbf{r} \mathrm{Re} \Sigma^R \cdot \nabla_\mathbf{k_\mathrm{F}} f + \\
&\qquad \qquad \qquad \qquad
\nabla_\mathbf{k_\mathrm{F}} \left( \epsilon_k + \mathrm{Re} \Sigma^R \right)
\cdot \nabla_\mathbf{r} f
= I_\mathrm{coll}.
\end{split}
\end{equation}
where $I_\mathrm{coll}$ and $\Sigma$ now contain the generalized distribution function 
$f(\mathbf{\hat{k}},\omega,\mathbf{r},t)$ instead of the associated Green's functions.  
We have introduced the notation $\nabla_\mathbf{k_\mathrm{F}} 
g(\mathbf{k},\omega)$ which is defined as $\nabla_\mathbf{k} g(\mathbf{k},\omega)$
evaluated at $\mathbf{k} = k_\mathrm{F} \mathbf{\hat{k}}$. 
Note that the term 
$\nabla_\mathbf{k_\mathrm{F}} \mathrm{Re} \Sigma^R \cdot \nabla_\mathbf{r} f$ is not zero
in general because the self-energy can still depend on $\mathbf{\hat{k}}$, however,
for this particular model the self-energy only depends on $\omega$ so this term can be dropped.

Eq. \ref{eq:fullQBE} is the full QBE for the generalized distribution function
$f(\mathbf{\hat{k}},\omega,\mathbf{r},t)$.  We see that despite the lack of 
a well-defined quasiparticle, this QBE looks very similar to the standard QBE
derived for Landau Fermi liquid theory.  Instead of the normal energy variable
$\epsilon_\mathbf{k}$, the QBE now contains $\omega$ which is independent
of $\mathbf{\hat{k}}$.  Moreover the QBE now involves renormalized
time and energy derivatives of the distribution function.\cite{Prange:64} 

We are interested in calculating the transport coefficients in the linear response regime.
In fact, we have already assumed that the deviations from equilibrium are small, so that the
generalized distribution functions are well-defined even out of equilibrium.  We therefore
proceed to linearize the QBE of Eq. \ref{eq:fullQBE}.  By linearizing the QBE in a particular
way, we are able to use a variational method to estimate the transport coefficients.
Focusing on the RHS of Eq. \ref{eq:fullQBE}, \textit{i.e.} the collision integral,
we introduce $\mathbf{k}'$ and $\omega'$ as the energy and momentum of the intermediate
spinon in the 1-loop self-energy diagram and find that the collision term,
after the integration over $\xi$ from Eq. \ref{eq:QBE2}, becomes
\begin{widetext}
\begin{equation}
\begin{split}
I_\mathrm{coll} = N(0) \int d\,\omega' & d\,\mathbf{\hat{k'}} d\nu d\,\mathbf{q} \,\, 
\mathrm{Im} D(\mathbf{q},\nu) 
\left|\frac{\mathbf{k'}\times\mathbf{\hat{q}}}{m}\right| ^2  
\delta(k_F\mathbf{\hat{k'}}-k_F\mathbf{\hat{k}}-\mathbf{q}) \times \\
\biggl\{ &\delta (\omega'-\omega-\nu) \left[ n_0(\nu)(1-f(\mathbf{\hat{k'}},\omega'))
f(\mathbf{\hat{k}},\omega) - (n_0(\nu)+1)(1-f(\mathbf{\hat{k}},\omega)) 
f(\mathbf{\hat{k'}},\omega')\right] + \\
&\delta (\omega'-\omega+\nu) \left[ (n_0(\nu)+1)(1-f(\mathbf{\hat{k'}},\omega'))
f(\mathbf{\hat{k}},\omega) - n_0(\nu) f(\mathbf{\hat{k'}},\omega') 
(1-f(\mathbf{\hat{k}},\omega))\right] \biggr\},
\end{split}
\end{equation}
\end{widetext}
where $N(0) = m / \pi$ is the density of states at the Fermi level for the up
and down spins combined.
Here we have rearranged the terms from $\Sigma^>$ and $\Sigma^<$ into the two
processes corresponding to absorbing and emitting a gauge boson of energy $\nu$.
Note that from Eq. \ref{eq:prop},
\begin{equation}
\mathrm{Im} D(\mathbf{q},\nu)=\frac{\gamma \nu q}{\gamma^2 \nu^2+\chi_D^2 q^6},
\label{eq:Improp}
\end{equation}
where $q = |\mathbf{q}|$.

We define $f(\mathbf{\hat{k}},\omega) = 
f_0(\omega) + \delta f(\mathbf{\hat{k}},\omega)$, where $f_0(\omega)$ is
some local equilibrium distribution,
\begin{equation}
f_0(\omega,\mathbf{r},t)  = \frac{1}{e^{\beta(\mathbf{r},t) \left(\omega-\mu(\mathbf{r},t)\right)} + 1}.
\label{eq:led}
\end{equation}
We re-iterate that all of the distribution functions both in and out of equilibrium are 
also functions of $\mathbf{r}$ and $t$.  In particular the 
local equilibrium distribution $f_0(\omega)$ can depend on
space and time through the local temperature, $\beta(\mathbf{r},t)$.
This local equilibrium solution
has the property that it sets the collision terms on the RHS of the QBE
to be zero.\cite{Kadanoff} Thus from detailed balance, we derive the relations
\begin{align}
\label{eq:db1}
n_0 ( 1 - f'_0) f_0 = (n_0+1) (1-f_0) f'_0 
\end{align}
and
\begin{align}
(n_0+1) (1-f'_0) f_0 = n_0 f'_0 (1-f_0),
\label{eq:db2}
\end{align}
for the $\delta (\omega'-\omega-\nu)$ and $\delta (\omega'-\omega+\nu)$
processes respectively.  Note that for notational convenience we have 
introduced the definitions
$n_0 \equiv n_0(\nu)$, $f_0 \equiv f_0(\omega)$ and $f'_0 \equiv f_0(\omega')$.
We expect that for a fermion system the deviation from 
equilibrium $\delta f(\mathbf{\hat{k}},\omega)$ is
sharply peaked around the Fermi surface, $w=\mu$.  Therefore, we introduce a new function
$\phi(\mathbf{\hat{k}},\omega)$ defined by 
\begin{align}
\label{eq:phidef1}
f&=f_0 - \phi \frac{\partial f_0}{\partial \omega} \\
&=f_0 + \phi \beta f_0 (1-f_0).
\label{eq:phidef}
\end{align}
Thus the function $\phi$ is much smoother than the original $\delta f$.
Writing the generalized distribution function in this way is also critical
in deriving the variational method that we use to calculate the transport
properties.

Since we know that the local equilibrium distributions set the collision 
terms to be zero, we can expand out the distribution functions 
and using Eqs. \ref{eq:db1} and \ref{eq:db2} find that
\begin{equation}
\begin{split}
\label{eq:lincoll}
I_\mathrm{coll} = \int d\,&\omega' d\,\mathbf{\hat{k'}} d\nu d\,\mathbf{q} 
\,\, \mathrm{Im} D(\mathbf{q},\nu)  \times \\
&\left|\frac{\mathbf{k'}\times\mathbf{\hat{q}}}{m}\right|^2   
\delta(k_F\mathbf{\hat{k'}}-k_F\mathbf{\hat{k}}-\mathbf{q}) \times \\
&\bigl\{ \delta(\omega'-\omega-\nu) \beta (\phi-\phi') n_0 f_0 (1-f'_0) + \\
&\,\,\,\delta(\omega'-\omega+\nu) \beta (\phi-\phi') n_0 (1-f_0) f'_0 \bigr\},
\end{split}
\end{equation}
where $\phi \equiv \phi(\mathbf{\hat{k}},\omega$) and $\phi' \equiv \phi(\mathbf{\hat{k'}},\omega')$.

\section{Finite Temperature}

Before proceeding to calculating the transport coefficients using the linearized QBE,
we must consider carefully what happens at finite temperatures in this system.  This
problem was addressed by Kim \textit{et al.} in Ref. \onlinecite{Kim:95}.  We explain their
argument here so that we can see how it effects our derivation of the QBE and later our
derivation of the transport coefficients.
While in principle the derivations in section II are valid for both zero
and finite temperatures, in this system we need to take special care at 
finite temperatures because at any finite temperature T, the self-energy, 
$\mathrm{Im} \Sigma_0^R(\mathbf{k},\omega)$, is divergent even in equilibrium.\cite{Lee:92}
This is an infrared divergence and since it arises when the small 
$\mathbf{q}$ and $\omega$ limits have been treated exactly in cannot be
alleviated through some sort of cutoff.
One can show in fact that this divergence is an artifact of gauge invariance in the system.

The divergence of the self-energy implies that the spectral weight $A(\mathbf{k},\omega)$ and
the Green's functions are technically not well defined at finite temperature, which 
seems to invalidate the basis of our derivation of the QBE. This problem can be addressed
however by carefully considering the source of this divergence. 
In particular, we see that the source of the divergence is the 
gauge field fluctuations where the energy carried
by the gauge field $\nu$ is such that $\nu < T$.  We proceed then by
breaking up the gauge field fluctuations into two pieces.  Following Kim,
we define $\mathbf{a}_+(\mathbf{q},\nu)$ to be the fluctuations for $\nu > T$ and
$\mathbf{a}_-(\mathbf{q},\nu)$ to be the fluctuations for $\nu < T$.  We then
treat the $\mathbf{a}_-$ field as a vector potential which corresponds to 
a static applied random magnetic field $\mathbf{b}_- = \nabla \times \mathbf{a}_-$. 
We only consider the dynamics of the $\mathbf{a}_+$ field.  
It is then understood that all equations need to be averaged over all possible
configurations of this random magnetic field $\mathbf{b}_-$.

Kim \textit{et al.} showed that after breaking up the gauge field one 
can regain the original QBE if the momentum $\mathbf{k}$ is shifted to 
$\mathbf{k}_-=\mathbf{k}-\mathbf{a}_-$ and 
the self-energy is understood to only include fluctuations of the $\mathbf{a}_+$
field.  Since the self-energy now contains only fluctuations with $\nu>T$, it is 
no longer divergent. One then recovers the original QBE of Eq. \ref{eq:fullQBEunint} 
with an additional term on the LHS
\begin{equation}
\frac{\mathbf{k_-}}{m} \cdot \mathbf{b}_- \times \nabla_\mathbf{k_-} G^<
\end{equation}
corresponding to an applied random field $\mathbf{b}_-$.  
The original divergence of the full self-energy can be understood as a consequence
of the non-gauge invariance of the original Green's function.  We note that the
extra term added to the QBE by breaking up the gauge fluctuations depends only 
on the gauge invariant quantity $\mathbf{b}_-$ and not on the potential $\mathbf{a}_-$.

We now proceed to consider the effect of this applied random field on transport
calculations using the linearized QBE that we have derived.  In principle, we
calculate the transport properties separately for each possible configuration
of the random field and then averages over them all.  Because of this averaging,
the field $\mathbf{b}_-$ cannot give rise to a linear response unlike an applied
``electric field'' or thermal gradient. It can however affect the transport coefficients 
of these quantities by contributing an additional source of scattering.  
We can estimate the effect of this scattering by calculating
a scattering rate due to $\mathbf{a}_-$ fluctuations.  Kim \textit{et al.}
show that these fluctuations give rise to a scattering rate
$\tau^{-1} \sim T^{4/3}$.  We can then use this estimate to calculate
how this scattering affects the low temperature forms of the transport
coefficients.

\section{Spin Resistivity}

The first transport property that we calculate is the spin resistivity.
While less experimentally accessible than the thermal conductivity, it is 
technically simpler and thus serves to illustrate our method.
We assume that there is some uniform applied force field $\mathbf{F}$ that 
couples linearly to the spinons.  One possible way of realizing such a field
is by applying a spatially varying magnetic field.  
Assuming that $B$ is uniform along one direction and has a constant slope along the
other direction, the spinons see a uniformly applied force field, 
$\mathbf{F} = \nabla B$. This field pushes up spins and down spins in 
opposite directions and thus leads to a net spin current, $\mathbf{J}_\mathrm{s}$.  

This effective applied force field $\mathbf{F}$ couples to the spinons just as
an electric field couples to an electron with unit charge.  Thus we can
borrow the result from a system of electrons in an applied electric field to
determine the form of the driving term in the QBE. Following the work of Mahan 
in Ref. \onlinecite{Mahan:83}, we find that in the presence of
a scalar potential, the energy of the particle depends on its location,
where as in steady state we expect the system to be spatially uniform.
Mahan shows that this problem can be eliminated through a change of variables which has the
secondary effect of changing the derivative $\nabla_\mathbf{r}$ to
$\nabla_\mathbf{r} + \mathbf{F} \frac{\partial}{\partial \omega}$.  This
change of variables generates the driving term in the LHS of the QBE.

We again linearize the QBE using the earlier definitions of
$f_0(\omega,\mathbf{r},t)$ and $\phi(\mathbf{\hat{k}},\omega,\mathbf{r},t)$.
Since the applied force field $\mathbf{F}$ is independent of position and time, 
we expect that that the steady state solutions for both $f_0(\omega)$ and 
$\phi(\mathbf{\hat{k}},\omega)$
are independent of $\mathbf{r}$ and $t$.  Thus returning to Eq. \ref{eq:fullQBE}, we see that 
the first two terms are zero.  
Using the spatial invariance of the distribution functions and 
the fact that the equilibrium distribution function is independent of $\mathbf{\hat{k}}$, 
we are left with
\begin{equation}
\label{eq:Fqbe}
-\mathbf{F} \cdot \mathbf{\hat{k}} v_\mathrm{F} 
\frac{\partial f_0}{\partial \omega} = I_\mathrm{coll}.
\end{equation}
This is the QBE that we consider for calculating the spin resistivity.  
Note that all of the terms containing the self-energy on the LHS have disappeared from
the QBE.  Thus the divergent effective mass does not even enter this calculation and
should have no effect on the spin resistivity. We also note that Eq. \ref{eq:Fqbe} 
is almost identical to the linearized Boltzmann equation in an applied electric field with
the change that $\mathbf{k}$ is now $(\mathbf{\hat{k}}, \omega$).  Thus we
proceed by following Ref. \onlinecite{Ziman} to derive a variational method
for calculating the transport coefficients. 

We begin with defining the LHS of the linearized QBE, here Eq. \ref{eq:Fqbe},
 to be $X(\mathbf{\hat{k}},\omega)$.
We then rewrite the collision term by defining a function 
$P(\mathbf{\hat{k}},\omega,\mathbf{\hat{k'}},\omega')$ that is the analogue
of the equilibrium
transition rate between the states $(\mathbf{\hat{k}},\omega)$ and 
$(\mathbf{\hat{k'}},\omega')$.
Therefore the linearized QBE of Eq. \ref{eq:Fqbe} can be written as
\begin{equation}
X(\mathbf{\hat{k}},\omega) = N(0) \int d\mathbf{\hat{k'}} d\omega'
\left(\phi-\phi'\right) P(\mathbf{\hat{k}},\omega,\mathbf{\hat{k'}},\omega').
\label{eq:formallinQBE}
\end{equation}

In order to proceed to derive the variational method we first define 
an inner product of functions of $(\mathbf{\hat{k}},\omega)$ given by
\begin{equation}
\left<g,h \right> \equiv N(0) \int d\mathbf{\hat{k}} d\omega \,\, 
g(\mathbf{\hat{k}},\omega) h(\mathbf{\hat{k}},\omega).
\end{equation}
We also define the operator $\mathcal{P}$ which takes the
function $\phi$ and returns the function given by
\begin{equation}
\mathcal{P} \phi \equiv N(0) \int d\mathbf{\hat{k'}} d\omega'
\left(\phi-\phi'\right) P(\mathbf{\hat{k}},\omega,\mathbf{\hat{k'}},\omega').
\label{eq:Pdef}
\end{equation}
We note that the equilibrium transitional rate $P$ is symmetric in the
exchange of $(\mathbf{\hat{k}},\omega)$ with $(\mathbf{\hat{k'}},\omega')$.
It is easy to see that the operator $\mathcal{P}$ is linear and 
$\left< \phi, \mathcal{P} \phi \right> \geq 0$.
From these properties, we can derive that the solution $\phi$ of any
linearized QBE that can be written in the form of Eq. \ref{eq:formallinQBE} is 
such that it minimizes the quantity, \cite{Ziman}
\begin{equation}
\Delta = \frac{\left<\phi,P\phi\right>}{\left< \phi,X \right>^2}.
\label{eq:minquant}
\end{equation}

We now return to consider the specific case of the QBE given by Eq. \ref{eq:Fqbe}.  
Since the up and down spins contribute identically to the spin current
$\mathbf{J}_\mathrm{s}$ and we are interested in $\rho_S$ only up to a numerical prefactor,
we can consider the current due to the motion of only one type of spin which we
denote by $\mathbf{J}$.
For a current density $\mathbf{J}$ the 
energy density dissipation rate is $\rho_\mathrm{s} \mathbf{J}^2$.  This can
in turn be related to the rate of entropy density production, so that
we are left with
\begin{equation}
T \dot{S} = \rho_\mathrm{s} \mathbf{J}^2,
\label{eq:Sdot}
\end{equation}

Looking at the definition of the spin current in terms of the 
generalized distribution function, we find that
\begin{equation}
\mathbf{J} = N(0) \int d\mathbf{\hat{k}} d\omega \, v_\mathrm{F} \mathbf{\hat{k}} 
f(\mathbf{\hat{k}},\omega).
\end{equation}
It is important to note that none of the renormalizations from the self-energy
enter this expression.  These renormalization factors instead appear on the
time derivative of the particle density.\cite{Prange:64}  This fact is critical for
relating the variational principle to the transport coefficients and also
helps explain why the diverging effective mass does make the
transport coefficient ill-defined.

Because of the $\mathbf{\hat{k}}$ independence of $f_0(\omega)$,
\begin{equation}
\mathbf{J} = - N(0) \int d\mathbf{\hat{k}} d\omega \, v_\mathrm{F} \mathbf{\hat{k}}
\phi(\mathbf{\hat{k}},\omega) \, \frac{\partial f_0(\omega)}{\partial \omega}.
\label{eq:spincurrent}
\end{equation}
Thus if $X$ is evaluated for unit applied force field, $|\mathbf{F}| = 1$,
we are left with
\begin{equation}
\left| \mathbf{J} \right|^2 = \left< \phi, X \right>^2.
\label{eq:denom}
\end{equation}

Following Ref. \onlinecite{Ziman}, we now consider how the entropy
is related to the distribution function $f$.
In general the Boltzmann equation balances the rate of
change of the distribution function due to diffusion, external fields and scattering.  
We can view the QBE of Eq. \ref{eq:fullQBE} in the same light.  
From statistical mechanics we know that the entropy of a system of fermions in
equilibrium is given by 
\begin{equation}
S = -k_\mathrm{B} N(0) \int \left[f \ln f + (1-f) \ln (1-f)\right] d \mathbf{\hat{k}} d \omega.
\end{equation}
We assume that this formula holds for small deviations around equilibrium.
Differentiating with respect to time and linearizing, the rate of entropy
production is given by
\begin{equation}
\dot{S} = - \frac{N(0)}{T} \int \phi \dot{f} d \mathbf{\hat{k}} d \omega.
\end{equation}
Thus from the collision term, we see that the rate of entropy production 
due to scattering is given by
\begin{equation}
T \dot{S} = \left< \phi, \mathcal{P} \phi \right>.
\label{eq:Sdoteq}
\end{equation}

In steady state the macroscopic rate of entropy production, 
\textit{i.e.} the Joule heating, is equal to the entropy production due to scattering.
Thus we can solve Eq. \ref{eq:Sdot} for $\rho_\mathrm{s}$ and using Eqs. \ref{eq:denom} and 
\ref{eq:Sdoteq}, we find
\begin{equation}
\rho_\mathrm{s} = \frac{ \left<\phi, \mathcal{P} \phi \right>}{\left< \phi, X \right>^2}.
\label{eq:rhoSeq}
\end{equation}
Comparing Eq. \ref{eq:rhoSeq} to the earlier definition of $\Delta$, we see that the
solution $\phi$ that solves the linearized QBE of Eq. \ref{eq:Fqbe} minimizes
$\rho_\mathrm{s}$.  We thus proceed using the standard variational method.
By making a reasonable ansatz for the deviation from local equilibrium $\phi$,
we can calculate an estimate for the spin resistivity $\rho_\mathrm{s}$.

Although the collision term includes the two distinct processes of absorbing and
emitting a gauge boson, we note that we need only calculate
$\rho_\mathrm{s}$ for one of the two processes since they contribute identically to
the spin resistivity.  We consider 
the trial function 
\begin{equation}
\phi(\mathbf{\hat{k}},\omega) = \eta \,( \mathbf{\hat{k}} \cdot \mathbf{\hat{F}}) 
\label{eq:resphidef}
\end{equation}
where $\mathbf{\hat{F}}$ is a unit vector in the direction of the applied field 
$\mathbf{F}$ and $\eta$, which has units of $k_\mathrm{F}/m$, is small.  
This particular deviation from equilibrium $\phi$ can be interpreted as
a shift of the Fermi surface in the $\mathbf{\hat{F}}$ direction
which is a reasonable guess since the shifted Fermi surface is the ansatz used to derive the
standard $T^5$ dependence of the low temperature resistivity in a metal.

Both the numerator and the denominator of Eq. \ref{eq:rhoSeq} contain factors of
$\eta^2$ so they cancel and we drop this factor from the rest of our
calculation.  Plugging in the particular form of the trial function $\phi$ from 
Eq. \ref{eq:resphidef} into Eq. \ref{eq:spincurrent},
it is straightforward to calculate the denominator which is 
independent of $T$ and given by
\begin{equation}
\left< \phi, X \right>^2 = \left(m v_\mathrm{F} \right)^2.
\label{eq:finaldenom1}
\end{equation}
   
Looking back at the definition of the operator $\mathcal{P}$ given in Eq. \ref{eq:Pdef} and
using its symmetry properties, the numerator of Eq. \ref{eq:rhoSeq}  
up to a numerical prefactor is given by
\begin{align}
\nonumber
\left< \phi, \mathcal{P} \phi \right> =  
\, \beta \int &d\omega d\omega' d\nu d\mathbf{q} d\mathbf{\hat{k}} 
d\mathbf{\hat{k'}} 
\left|\mathbf{k'}\times\mathbf{\hat{q}}\right|^2 \times \\
&\mathrm{Im} D(\mathbf{q}, \nu) 
\nonumber (\phi-\phi')^2  f_0(\omega) (1-f_0(\omega')) \times \\
&n_0(\nu) \delta(\omega'-\omega-\nu) \delta(\mathbf{\hat{k'}}-\mathbf{\hat{k}} 
- \frac{\mathbf{q}}{k_\mathrm{F}}).
\label{eq:Entropy}
\end{align}
We first perform the integration over $\omega$ and $\omega'$, using the result that
\begin{align}
\nonumber \int \,d\omega\,d\omega' \delta(\omega'-\omega-\nu) 
f_0(\omega)&(1-f_0(\omega')) \\
&= \frac{z}{\beta(1-e^{-z})},
\end{align}
where $z=\beta \nu$.
We define $\theta$ and $\theta'$ as the angles between
$\mathbf{\hat{k}}$ and $\mathbf{\hat{u}}$, and $\mathbf{\hat{k'}}$ and 
$\mathbf{\hat{u}}$ respectively.  We then define $\alpha = \theta'-\theta$.
The integration over $\mathbf{q}$ just 
enforces the condition from the momentum delta function, so that we are left with
\begin{align}
\nonumber
\left< \phi, \mathcal{P} \phi \right> 
= \beta &\int d\nu d\theta d\theta' 
\frac{m^2 v_\mathrm{F}^2}{2} \frac{\sin^2(\alpha)}{1-\cos(\alpha)} 
\mathrm{Im} D(q(\alpha),\nu) \times \\ 
&(\cos\theta - \cos\theta')^2 \frac{z}{\beta(1-e^{-z})(e^{z}-1)},
\end{align}
where we have defined the function $q(\alpha)$ which gives the magnitude
of $q$ for a particular angle $\alpha$.
We shift the integration over $\theta$ and $\theta'$ to $\alpha =
\theta'-\theta$ and $\alpha'=\theta'+\theta$.  Integrating over
$\alpha'$ gives
\begin{align}
\nonumber
\left< \phi, \mathcal{P} \phi \right> 
= &\beta \int  d\nu d\alpha \, m^2 v_\mathrm{F}^2 \frac{\sin^2(\alpha)}{1-\cos(\alpha)} 
\mathrm{Im} D(q(\alpha),\nu) \times \\
&\sin^2(\frac{\alpha}{2})
(2\pi-\alpha+\sin \alpha) \frac{z}{\beta(1-e^{-z})(e^{z}-1)}.
\end{align}
We now assume that since $\mathrm{Im} \, D(q,\nu)$ is peaked 
for small $q \ll k_\mathrm{F}$ that we can take $\alpha$ to be small.
Re-introducing $q$ through a change of variables of $\alpha$, we find
\begin{align}
\left< \phi, \mathcal{P} \phi \right> 
= \beta \int d\nu dq \frac{q^2}{k_\mathrm{F}} \mathrm{Im} D(q,\nu) \frac{z}{\beta(1-e^{-z})(e^{z}-1)}.
\end{align}

Inserting the RPA propagator from Eq. \ref{eq:Improp}, we arrive at
\begin{align}
\nonumber
\left< \phi, \mathcal{P} \phi \right> = 
\left(\frac{\gamma^{1/3}}{\chi_D^{4/3} k_\mathrm{F} \beta^{4/3}} \right)
&\left( \int_1^\infty 
dz \frac{z^{4/3}}{\left(1-e^{-z} \right) \left(e^z-1 \right)} \right) \times \\
&\,\,\,\,\,\, \left( \int_0^{ y(q=2 k_\mathrm{F})} d\,y \frac{y^{1/3}}{1+y^2}\right).
\label{eq:finalnum1}
\end{align}
The lower limit of the integration over $z$ is an artifact of our treatment
of the divergence of the self-energy at finite temperatures as noted in section III.
The self-energies in the QBE are taken to only include gauge fluctuations with 
$\nu > k_\mathrm{B} T$.
We consider the effect of the low energy gauge fluctuations below.

Looking at the integrand of the $z$ integral in Eq. \ref{eq:finalnum1}, we see that 
it is sharply peaked for small
$z$.  Since $y \sim 1/z$ we can thus take the limit of the integration over $y$
to be $\infty$.  The integrals over $y$ and $z$ are thus numerical prefactors of order
unity that we ignore.  Combining Eqs. \ref{eq:rhoSeq}, \ref{eq:finaldenom1} 
and \ref{eq:finalnum1}, and reinserting the correct factors of $\hbar$,
we find that the spin resistivity is, up to a constant of
order one,
\begin{equation}
\rho_\mathrm{s} = \hbar \left( \frac{k_\mathrm{B} T} {\epsilon_\mathrm{F}}\right)^{4/3}.
\end{equation}
The final result for the spin resistivity takes exactly the form predicted by naive arguments 
assuming the existence of quasiparticles and ignoring the effects of the mass renormalization.
Thus the only effect of the gauge bosons is through the relaxation rate.
However, we have now derived this result without assuming the existence of quasiparticles and 
we have shown that a possible divergence due to the effective mass does not enter the 
expression for the spin resistivity.

Finally, we briefly discuss the effect of the low energy gauge fluctuations that cause the
divergence of the self-energy at finite temperatures.  As described in section III, these
fluctuations enter the QBE as an applied static random magnetic field $\mathbf{b}_-$.
Averaging over all possible configurations of this field does not lead to a linear
response.  Its only effect is through the transport scattering rate $\tau_- \sim T^{-4/3}$.   
Taking the result from the QBE derivation, we assume that we can use the naive form for the
spin resistivity ignoring the mass renormalization.  Thus we expect that the low
energy fluctuations in the self-energy loop do not change the overall scaling of
$\rho_\mathrm{s}$ and only enter through the numerical prefactor.

\section{Thermal Conductivity}

We now proceed to calculate the more experimentally relevant transport coefficient,
the thermal conductivity.  We consider the situation where a uniform thermal gradient 
is applied to the system giving rise to a heat current density 
$\mathbf{U}$ which is related to the thermal gradient through $\mathbf{U} = \kappa \nabla_\mathbf{r} T$
where $\kappa$ is the thermal conductivity.  The heat current density $\mathbf{U}$ 
is defined in terms of the energy current density $\mathbf{J}_\mathrm{E}$ and 
the particle current density $\mathbf{J}_\mathrm{p}$, where
\begin{equation}
\mathbf{U} = \mathbf{J}_\mathrm{E} - \mu \mathbf{J}_\mathrm{p}.
\end{equation}
Note that the particle current is not equivalent to the spin current $\mathbf{J}_\mathrm{s}$.
Despite the fact that the quasiparticle is not well defined, these currents can be
related to the generalized spinon distribution function $f(\mathbf{\hat{k}},\omega)$ through 
\begin{align}
&\mathbf{J}_\mathrm{E} = N(0) \sum_\sigma \int \omega v_\mathrm{F} \mathbf{\hat{k}} f_\sigma
(\omega,\mathbf{\hat{k}}) d\mathbf{\hat{k}} d\omega \\
&\mathbf{J}_\mathrm{p} = N(0) \sum_\sigma \int v_\mathrm{F} \mathbf{\hat{k}} f_\sigma
(\omega,\mathbf{\hat{k}}) d\mathbf{\hat{k}} d\omega.
\end{align}
As in the case of the spin current, we again see that the renormalizations due to the 
self-energy do not enter the expressions for the currents and instead appear on the
associated time derivatives of density or energy.\cite{Prange:64}

Thus the heat current density $\mathbf{U}$ is given by
\begin{align}
\mathbf{U} = N(0) \sum_\sigma \int v_\mathrm{F} \mathbf{\hat{k}} (\omega - \mu) 
f_\sigma (\omega,\mathbf{\hat{k}})
d\mathbf{\hat{k}} d\omega.
\label{eq:heatcurrent}
\end{align}
In the case of an applied thermal gradient, spin up and spin down react
identically so we can drop the spin index for the remainder of the calculation.

For a system in steady state in an applied thermal gradient, its clear that
the local equilibrium $f_0(\omega)$ defined in Eq. \ref{eq:led} depends
on position $\mathbf{r}$ through a local temperature $\beta(\mathbf{r})^{-1}$.
The distribution function however remains time independent.  
Thus the QBE for the system under these conditions is 
\begin{equation}
- \nabla_\mathbf{r} \mathrm{Re} \Sigma^R \cdot \nabla_\mathbf{k_\mathrm{F}} f + 
\nabla_\mathbf{k_\mathrm{F}} \epsilon_k \cdot \nabla_\mathbf{r} f
 = I_\mathrm{coll}.
\label{eq:thermqbe}
\end{equation}

We again linearize, assuming that $f$ is expanded into $f_0$ and $\phi$ as
defined in Eq. \ref{eq:phidef} and that $\nabla T / T$ is small.
Because $f_0(\omega)$ is assumed
to be $\mathbf{\hat{k}}$ independent, the linearized QBE becomes
\begin{equation}
\nabla_\mathbf{k_\mathrm{F}} \epsilon_k \cdot \nabla_\mathbf{r} f_0
 = I_\mathrm{coll}.
\label{eq:linthermqbe}
\end{equation}

The earlier derivation of the variational principle remains true for Eq. 
\ref{eq:linthermqbe} as well.  Thus the quantity $\Delta$, 
defined in Eq. \ref{eq:minquant} is still minimized by the solution $\phi$ 
of the linearized QBE.  Note that $\Delta$ is dependent on the particular
QBE that we are considering. To proceed, we
relate $\Delta$ for the linearized QBE of Eq. \ref{eq:linthermqbe}
to the thermal conductivity and then, in an identical
way to the spin conductivity calculation, guess a reasonable $\phi$ 
and calculate the associated thermal conductivity $\kappa$.

We again start with an expression for the rate of entropy density production.\cite{Ziman}
For a given thermal current density, we have an entropy flux given by
$\mathbf{U}/T$.  This leads to a rate of entropy density production 
\begin{equation}
\dot{S} = \nabla(\frac{1}{T}) \cdot \mathbf{U} = \frac{U^2}{\kappa T^2},
\end{equation}
since $\mathbf{U}$ is constant throughout the sample.  Note that we are
assuming that there is no particle current due to the applied thermal gradient and
thus no entropy production associated with a dissipation due to the presence of 
such a current.  This assumption is valid for this system as long as there is no source of 
spinons.  To see this we return to the Lagrangian of Eq. \ref{lagrangian}.
Integrating out the gauge field $\mathbf{a}$ leads to the constraint that the spinon particle current 
density is conserved.  Thus, the thermal resistivity $W = \kappa^{-1}$ is given by
\begin{equation}
W = \frac{T^2 \dot{S}}{U^2}.
\label{eq:Wdef}
\end{equation}

The derivation of $\dot{S}$ in terms of $\phi$ and $\mathcal{P} \phi$ from
Eq. \ref{eq:Sdoteq} is still valid here.  Thus we only need to consider 
$\left< \phi, X \right>$ and see if it is related to 
the heat current density $\mathbf{U}$.
We again define the LHS of the linearized QBE, now Eq. \ref{eq:linthermqbe},
to be $X$.  Thus,
\begin{align}
\left< \phi, X \right> &= N(0) \int d\mathbf{\hat{k}} d\omega
\phi v_\mathrm{F} \mathbf{\hat{k}} \nabla_\mathbf{r} f_0 \\
&= - N(0) \int d\mathbf{\hat{k}} d\omega \phi v_\mathrm{F}
\beta (\omega -\mu) \frac{\partial f_0}{\partial \omega}
\nabla_\mathbf{r} T \cdot \mathbf{\hat{k}}.
\end{align}
Inserting the expression for $f$ from Eq. \ref{eq:phidef1} into the expression for the
heat current density $\mathbf{U}$ from Eq. \ref{eq:heatcurrent}, gives an identical 
expression up to a factor of $\beta$.  We find that
\begin{equation}
\left< \phi, X\right> = - \beta \, \nabla_\mathbf{r} T \cdot \mathbf{U},
\end{equation}
and $\mathbf{U}^2$ is thus equal to the denominator of $\Delta$
up to some temperature independent constant.  Therefore the $\phi$ that solves the linearized
QBE of Eq. \ref{eq:linthermqbe} minimizes the thermal resistivity $W$.  Given an arbitrary $\phi$,
$W$ can be calculated, up to some numerical prefactor, with 
\begin{equation}
W = \frac{T \left< \phi, \mathcal{P} \phi \right>}{ \left< \phi, X \right>^2}.
\label{eq:Wdef2}
\end{equation}

We need a trial function $\phi$ that leads to a net flow of heat but no spinon particle
current.  A trial function that almost realizes this condition is
\begin{equation}
\phi = \eta \, (\omega-\mu) \mathbf{\hat{k}} \cdot \mathbf{\hat{u}},
\label{eq:thermphidef}
\end{equation}
where $\mathbf{\hat{u}}$ is a unit vector in the direction of the heat current 
and $\eta$ is small.  This trial function can be interpreted as taking the 
original local temperature $\beta(\mathbf{r})$ and giving it a $\mathbf{\hat{k}}$ 
dependence, $\beta(\mathbf{\hat{k}}) = \beta + \eta \mathbf{\hat{k}} \cdot
\mathbf{\hat{u}}$.  Physically at some point $\mathbf{r}$, the spinons with momentum pointing in the
direction that the temperature decreases are hotter than average 
and on the opposite side of the Fermi surface the
spinons are colder than average.

With the trial function $\phi$ we now proceed to calculate the thermal resistivity $W$.  
First it is again clear that $W$ has no $\eta$ dependence.  Inserting the definition of the 
trial function $\phi$ from Eq. \ref{eq:thermphidef}, 
we find that $|\mathbf{U}| \sim m v_\mathrm{F} (k_\mathrm{B} T)^2$.
In order to calculate $\left< \phi, \mathcal{P} \phi \right>$, we must recalculate $(\phi-\phi')^2$ 
for the new trial function.  Using the relations imposed
by the energy and momentum delta functions, we find
\begin{align}
\nonumber
(\phi-\phi')^2 =& \nu^2 (\mathbf{\hat{k}} \cdot \mathbf{\hat{u}})^2 + (\omega'-\mu)^2
(\mathbf{q}/k_\mathrm{F} \cdot \mathbf{\hat{u}})^2  + \\ &2 \nu (\omega - \mu)
(\mathbf{\hat{k}} \cdot \mathbf{\hat{u}}) (\mathbf{q}/k_\mathrm{F} \cdot \mathbf{\hat{u}}).
\label{eq:Wnum}
\end{align}

Each of these terms can now be integrated individually in an analogous way
to the calculation of the numerator of the spin resistivity.
It is again necessary to use the fact that $\mathrm{Im} D(\mathbf{q},\nu)$ 
is peaked for small $q$ in order to arrive at an analytic result.
In evaluating these integrals, we need to calculate integrals of the form
\begin{equation}
I(n) = \int \,d\omega\,d\omega' \delta(\omega'-\omega-\nu) f_0(\omega)(1-f_0(\omega'))
(\omega -\mu)^n,
\end{equation}
for $n=0,1,2$.  The integrals can be evaluated and we find 
\begin{align}
&I(0)=\frac{z}{\beta(1-e^{-z})}, \\
&I(1)=\frac{-z^2}{2 \beta^2 (1-e^{-z})}, \\
&I(2)=\frac{\pi^2 z + z^3}{3 \beta^3 (1-e^{-z})}.
\end{align}

Using these results, we can now calculate the integrals, 
$\left<\phi, \mathcal{P} \phi \right>$ for each of the 
three terms in Eq. \ref{eq:Wnum}.  Dropping numerical prefactors, the
first term in Eq. \ref{eq:Wnum}, \textit{i.e.} the $\nu^2$ term,
contributes $v_\mathrm{F}^2 \left(m k_\mathrm{F}\right)^{2/3} (k_\mathrm{B} T)^{8/3}$ while
the remaining two terms both contribute $m^{4/3} k_\mathrm{F}^{-2/3} (k_\mathrm{B} T)^{10/3}$.
Combining this result with our earlier result for 
$\left<\phi,X\right>^2$ and 
plugging them into the definition of $W$ from Eq. \ref{eq:Wdef2}, 
we find that, restoring the correct factors of $\hbar$,
\begin{align}
W = \frac{\hbar}{k_\mathrm{B} \epsilon_\mathrm{F}}
\left[ \left( \frac{\epsilon_\mathrm{F}}{k_\mathrm{B} T}\right)^{1/3} +
\left(\frac{k_\mathrm{B} T}{\epsilon_\mathrm{F}}\right)^{1/3} \right] .
\end{align}
Thus to leading order the low temperature thermal conductivity $\kappa$ per layer is 
\begin{equation}
\frac{\kappa}{T} = \frac{k_\mathrm{B}^2}{\hbar} \left(\frac{\epsilon_\mathrm{F}}{k_\mathrm{B} T}\right)^{2/3}.
\label{eq:kappa}
\end{equation}
Thus we see that the QBE result for the thermal conductivity agrees with the result we found
by making the unjustified assumption that quasiparticles were well-defined and that we can ignore
the effects of mass renormalization.  Note that these are the same assumptions that gave the correct
result for the spinon resistivity.
 
With these assumptions we can show that $\kappa / T \sim \tau_E$. We now use this
result to consider the effects of the low energy $\mathbf{a}_-$ fluctuations.
The applied static random field $\mathbf{b}_-$ again only enters the thermal
conductivity through a relaxation rate since it cannot lead to a linear response.
We therefore are interested in the contribution to the energy relaxation rate 
due to $\mathbf{b}_-$ fluctuations. This contribution goes as $T^{-2/3}$.  
Thus if we assume that the QBE result that $\kappa / T \sim \tau_E$ is valid ,
the low energy gauge fluctuations only change the numerical prefactor of $\kappa$ 
and not its temperature dependence.

The form of the thermal conductivity from Eq. \ref{eq:kappa} is valid for clean
systems only.  In reality the divergent behavior of $\kappa / T$ at low $T$
will be cutoff by the impurity scattering rate $1/\tau_0$.  Again assuming
that the QBE result is valid we thus find that the overall result for the thermal
conductivity per layer in the presence of impurities is
\begin{equation}
\frac{\kappa}{T} \sim \left( \frac{\hbar}{k^2_\mathrm{B}} 
\left( \frac{k_\mathrm{B} T}{\epsilon_\mathrm{F}} \right)^{2/3} + 
\frac{m A}{k_\mathrm{B}^2}\frac{1}{\tau_0} \right)^{-1},
\end{equation} 
where $A$ is the area of the layer.

\section{Thermal Conductivity of Gauge Bosons}

In all of the calculations up to this point we have assumed that the gauge bosons
are always in local thermal equilibrium.  Thus when we wrote down the QBE for the 
spinons, we assumed there was no deviation from the standard form for $n_0(\nu)$.
In this section we examine the validity of this approximation
and in particular look at the corrections to the thermal conductivity 
that arise when this assumption is relaxed.  

As mentioned in the introduction, in the simplest derivation of the thermal conductivity, 
one finds that $\kappa$ is proportional to the heat capacity of the particles.
It is known that for the model of Eqs. \ref{lagrangian} and \ref{eq:prop}
the abundance of soft gauge field fluctuations produce a $T^{2/3}$ contribution 
to the specific heat.\cite{Motrunich:05}  As we showed in the previous section
the effective mass drops out of the expression for the thermal conductivity
due to the spinons.  Ignoring the mass renormalization, the specific heat from the
spinons is linear in $T$.  Therefore, it is possible that at low temperatures, 
the largest contribution to the thermal conductivity comes from the gauge bosons. 
We now consider the effect of allowing them to deviate from 
thermal equilibrium.  This effect is the equivalent of phonon-drag in 
the electron-phonon system.

In order to describe the gauge bosons when they are not in thermal 
equilibrium, we derive the equation of motion for the gauge boson propagator that is analogous
to Eq. \ref{eq:QBE1}.  We then define a boson distribution function and derive
an associated QBE.  Thus we are left with two coupled differential equations
that describe the system.

Exactly as in section II, we begin by defining a number of different gauge boson propagators,
in analogy to the spinon propagators in Eqs. \ref{eq:Gdef} through \ref{eq:GRGA}
except with $\psi$ replaced by $a$.  We denote these propagators with $D$ rather than $G$.
Note that this is consistent with the definition from Eq. \ref{eq:prop} which we 
now associate with $D^R$.  We also
define an associated self-energy $\Pi(\mathbf{q},\nu)$ which arises from the spinon bubble.
In order to avoid double counting we consider the integration that lead to the gauge propagator
in a renormalization group sense.  In other words, we consider integrating out the high energy
spinons to generate the gauge propagator and then consider coupling of the gauge propagator
back to the low energy fermionic modes.  

Following the steps of section II, we consider the expression
\begin{align}
\nonumber D^>(\mathbf{q},\nu) - D^<(\mathbf{q},\nu) &= -2 \, \mathrm{Im} D^R \\
&= \frac{ -2 \gamma \nu q}{\gamma^2 \nu^2 + \chi_\mathrm{D}^2 q^6}.
\end{align}
It is clear that this expression is peaked at $\nu = \nu_\mathbf{q} = 
(\chi_\mathrm{D} / \gamma) q^3$.  We can integrate $\nu$ over the region of
peaking and define the gauge boson distribution function as
\begin{align}
n(\mathbf{q},\mathbf{r},t) = \int \frac{d\nu}{2\pi} D^<(\mathbf{q},\nu,\mathbf{r},t).
\end{align}
Note that unlike in the derivation of the generalized spinon distribution function, here 
we integrate over energy just as one does in the standard Fermi liquid case.

We now extended the assumption of the peaking of $D$ to situations near 
thermal equilibrium as well.  We see that despite the over-damped mode, the system looks
as if it has quasiparticles at $\nu = \nu_\mathbf{q}$.  This leaves us with the QBE
for the gauge boson distribution function, analagous to Eq. \ref{eq:fullQBE},
described by
\begin{align}
\frac{\partial n }{\partial t} + \nabla_\mathbf{q} \nu_q \cdot \nabla_\mathbf{r} n 
= \Pi^< (n+1) - \Pi^> n.
\label{eq:gaugeQBE}
\end{align}

We can now consider the two coupled QBEs of Eqs. \ref{eq:fullQBE} and \ref{eq:gaugeQBE}
to calculate the thermal conductivity.  Before proceeding, we linearize the gauge
boson QBE.  In analogy to Eq. \ref{eq:phidef}, we define the deviation from equilibrium to be 
\begin{align}
n(\mathbf{q}) &= n_0(\nu) - \zeta(\mathbf{q})
\frac{\partial n_0(\nu)}{\partial \nu} \\
&= n_0(\nu) + \beta \, \zeta(\mathbf{q}) n_0(\nu) \left( 1 + n_0(\nu) \right).
\end{align}
Note that in the above expressions $\nu$ is technically $\nu(\mathbf{q})$ since
we have defined the gauge boson distribution function as $n(\mathbf{q},\mathbf{r},t)$.

In principle, we would need
to solve the coupled set of QBEs given by Eqs. \ref{eq:fullQBE} and \ref{eq:gaugeQBE}.
However, in the case of thermal conductivity there is an important simplification.  We
can calculate the thermal conductivity due to the spinons and gauge bosons independently,
assuming in each case that the other type of excitations remain in thermal equilibrium.
This is valid because in this assumption, the terms that are ignored can be shown to be smaller
by a factor of $k_\mathrm{B} T / \epsilon_\mathrm{F}$ which is small for the temperatures
under consideration in this system.\cite{Ziman}

Thus to proceed we only need to calculate the additional contribution to the thermal conductivity
from the gauge bosons scattering off of spinons that can be assumed to be in thermal equilibrium.
With this assumption one can evaluate $\Pi^<$ and $\Pi^<$.  Remembering that the gauge propagator
$D(\mathbf{q},\nu)$ of Eq. \ref{eq:prop} was derived from the spinon bubble, we find that
after linearizing, $I_\mathrm{coll}$ for the boson QBE is given by
\begin{equation}
\begin{split}
I_\mathrm{coll} = \int d&\omega' d\,\mathbf{\hat{k'}} d\omega d\,\mathbf{\hat{k}} 
\,\, \mathrm{Im} D(\mathbf{q},\omega'-\omega) \beta \zeta(\mathbf{q}) \times \\
&n_0 (1-f_0) f'_0 \delta(k_F\mathbf{\hat{k'}}-k_F\mathbf{\hat{k}}-\mathbf{q}).
\end{split}
\end{equation}

We now make the ansatz $\zeta(\mathbf{q}) = \eta \mathbf{q} \cdot \mathbf{\hat{u}}$.  
Exactly as before one can derive an expression for the thermal conductivity in terms of a
variational method where
\begin{align}
\frac{1}{\kappa_g} = \frac{T^2 \dot{S}}{\mathbf{U}^2}.
\label{eq:kappag}
\end{align}
Just as in section V, the numerator can be related to the collision 
integral using Eq. \ref{eq:Sdoteq}, with $\phi(\mathbf{\hat{k}},\omega)$ 
replaced by $\zeta(\mathbf{q})$ and $\mathcal{P}$ now defined based on 
the gauge boson QBE.
In order to use the variational method, we also must consider the left-hand side
of the linearized form of the gauge boson QBE in Eq. \ref{eq:gaugeQBE} and 
relate it to $\mathbf{U}$.
\begin{align}
\left< \zeta, X \right> &= \int d\mathbf{q} \zeta(\mathbf{q}) \mathbf{v}_\mathbf{q} \cdot 
\nabla_\mathbf{r} n_0(\nu_\mathbf{q})  \\
&= \int d\mathbf{q}  \zeta(\mathbf{q}) \left(\mathbf{v}_\mathbf{q} \cdot \nabla T\right) 
\frac{\nu}{T} n(\mathbf{q}) \\
&= \frac{\nabla T}{T} \cdot \mathbf{U}
\end{align}
since for the gauge bosons
\begin{equation}
\mathbf{U} = \int d\mathbf{q} n(\mathbf{q}) \nu_\mathbf{q} \mathbf{v}_\mathbf{q}.
\end{equation}
Therefore we can use the variational method to calculate the gauge boson 
contribution to the thermal conductivity.

Returning to the denominator of Eq. \ref{eq:kappag}, we see that 
\begin{equation}
\left< \zeta(\mathbf{q}), X \right>^2 \sim T^2 \left| \int d\mathbf{q} \zeta(\mathbf{q}) 
\mathbf{v}_\mathbf{q} \frac{\partial n_0}{\partial T} \right|^2.
\end{equation}
Plugging in our variational ansatz for $\zeta(\mathbf{q})$, one can show 
that
\begin{equation}
\left< \zeta(\mathbf{q}), X \right>^2 \sim T^2 C^2
\end{equation}
where, as mentioned above the specific heat due to the gauge bosons $C \sim T^{2/3}$.

We now plug in the integral form of $\left< \zeta, \mathcal{P}\zeta \right>$
into Eq. \ref{eq:kappag} giving
\begin{align}
\nonumber \frac{1}{\kappa_g} \sim \frac{1}{T C^2} 
\int d\omega &d\omega' d\mathbf{q} d\mathbf{\hat{k}} d\mathbf{\hat{k'}} 
\mathrm{Im} D(\mathbf{q}, \omega'-\omega) \beta \zeta^2(\mathbf{q}) \times \\
&n_0 (1-f_0) f'_0 \delta(k_F\mathbf{\hat{k'}}-k_F\mathbf{\hat{k}}-\mathbf{q}).
\end{align}
For the particular ansatz $\zeta$ that we are considering, this integral
is nearly identical to the calculation of the spin resistivity.  After some
work and borrowing the results from earlier calculations in this paper, we find
\begin{equation}
\frac{1}{\kappa_g} \sim \frac{1}{T C^2} T^{4/3} \sim \frac{1}{T}.
\end{equation}
Therefore the total thermal conductivity is
\begin{equation}
\kappa \sim c_s T^{1/3} + c_g T.
\end{equation} 
At low temperatures the thermal conductivity is thus dominated by the 
contribution from the spinons and one can ignore the equivalent of 
the phonon-drag.

\section{Conclusion}

Motivated by the organic compound $\kappa$-(BEDT-TTF)$_2$-Cu$_2$(CN)$_3$ 
which is a quasi two-dimensional effectively isotropic spin $1/2$ 
Heisenberg model on the triangular lattice, we studied a particular $U(1)$ 
spin liquid defined by Eqs. \ref{lagrangian} and \ref{eq:prop}, that has
been recently proposed to describe this system at low temperatures.  From this
model, a variety of properties that can be compared to experiment 
such as the specific heat and static spin susceptibility have been calculated.
In this paper we have continued along these lines by calculating the
transport properties of this spin liquid.

In addition to the physical relevance of this calculation, it is also
interesting from a theoretical point of view because of problems 
that arise in deriving these coefficients.  In particular, there are three main
issues that we had to circumvent.  First, the renormalized spinons are not
well-defined Landau quasiparticles because the one-loop correction to the
spinon self-energy scales as $\omega^{2/3}$.  Second, the effective mass 
diverges at the Fermi surface.  And finally, as an artifact of gauge-invariance, 
the self-energy is divergent at finite temperature.  These issues clearly
invalidate the results one finds from naive arguments using renormalized
parameters.

Despite these problems, we were able to proceed by deriving a QBE for this system.
Starting from Dyson's equation, we showed that even though there is no well-defined 
concept of a quasiparticle, we can construct a generalized distribution 
function based on the peaking of the spectral weight as a function of $\xi_\mathbf{k}$, 
instead of the standard $\omega$ peaking.  Using this peaking, we derived the QBE for 
this system in terms of the generalized distribution function.

We then proceeded to linearize the spinon QBE and to calculate the transport properties.
Applying the variational method for solving the Boltzmann equation to the QBE, we
solved for the transport coefficients.  In particular, we showed that the transport
coefficients are well-defined despite the diverging effective mass.  Moreover,
we calculated the temperature dependence of the spin resistivity and thermal conductivity.
Assuming the gauge bosons remain in thermal equilibrium, we found that at low temperatures,
the spin resistivity goes as $T^{4/3}$ and the thermal conductivity goes as $T^{1/3}$.
In both cases, the result is identical to the one derived from naive arguments,
where all the renormalizations due to the one loop corrections are ignored and the only
effect of spinon-boson interaction is through a scattering rate $1/\tau$.

In the final section of this paper, we relaxed the assumption that the gauge bosons are in 
thermal equilibrium.  We showed that one can follow a similar method to the derivation of
the spinon QBE and derive a QBE for the gauge bosons.  In principle, we would then need
to solve a coupled set of differential equations; however, for the specific case
of the thermal conductivity we can decouple the equations if we ignore terms of order 
$k_\mathrm{B} T / \epsilon_\mathrm{F}$.  Under this assumption we showed that
the gauge bosons contribution to the thermal conductivity is sub-dominant to the
spinon contribution.

Finally, we comment on the specific measurement of these transport coefficients in 
$\kappa$-(BEDT-TTF)$_2$-Cu$_2$(CN)$_3$.  As mentioned in the beginning of section III,
the spin resistivity is difficult to measure because of the problem of
creating a source and drain of spinons.  The thermal conductivity however is experimentally
accessible.  Recent experimental and theoretical work on the specific heat and spin
susceptibility have show evidence of spinon pairing at low temperatures (around 6K) in this system.
Clearly the results in this work are invalid in the pairing regime; however, the exchange
coupling is estimated to be $J \sim 250 K$, so there is a wide region of temperatures
where one could see evidence of the spin liquid state through the temperature dependence of 
the thermal conductivity.

We thank Y. B. Kim and  K. Kanoda for their discussions and help on this research.
P.A.L acknowledges support by NSF grant DMR-0517222.

\bibliography{tjbib}

\end{document}